%Paper: hep-ph/9408272
%From: GOLDEN@huhept.harvard.edu
%Date: Thu, 11 Aug 1994 13:05:19 -0400 (EDT)

% this file uses harvmac for the text and pictex for the figure.  Both of
% these macro packages may be obtained from xxx.lanl.gov.  If you don't have
% pictex on your system, just answer "n" to the question, and the figure will
% be omitted.

\def\yesans{y }
\message{ Do you have the hypertex macros, lanlmac.tex, installed on your
system (y/n)?}\read-1 to\ansm
\ifx\ansm\yesans\message{Using hypertex macros.}
\input lanlmac

\else\message{Too bad!  Using harvmac instead.}
\input harvmac

\fi

\message{ Do you have pictex installed on your system (y/n)? }\read-1 to\ansp
\ifx\ansp\yesans\message{(The figure(s) will be included.)}
\def\inpic{\input pictex}
\else\message{(The figure(s) will be omitted.)}
\def\inpic{\end}
\fi

\noblackbox
\def\pr#1#2#3{Phys. Rev. {\bf #1}, #2 (#3)}
\def\prl#1#2#3{Phys. Rev. Lett. {\bf #1}, #2 (#3)}
\def\plb#1#2#3{Phys. Lett. B {\bf #1}, #2 (#3)}
\def\prd#1#2#3{Phys. Rev. D {\bf #1}, #2 (#3)}
\def\npb#1#2#3{Nucl. Phys. B {\bf #1}, #2 (#3)}
\def\mpla#1#2#3{Mod. Phys. Lett. A {\bf #1}, #2 (#3)}
\def\physica#1#2#3{Physica {\bf #1}, #2 (#3)}
\def\tr{\hbox{tr }}

\Title{\vbox{\baselineskip12pt\hbox{HUTP-94/A020}
\hbox{hep-ph/9408272}}}
{\vbox{
\centerline{Unitarity and Fermion Mass Generation}
}}

\centerline{Mitchell Golden$^*$}
\footnote{}{$^*$Lyman Laboratory of Physics,
Harvard University, Cambridge, MA 02138}
\footnote{}{golden@physics.harvard.edu}

\vskip .4in

\centerline{\bf ABSTRACT}

Some years ago Appelquist and Chanowitz considered the scattering of
fermion--anti-fermion into a pair of longitudinal gauge bosons.  Their
calculation established that unitarity implies that the physics giving mass to
a quark of mass $m_f$ must be below a mass scale of $16 \pi v^2 / m_f$
($v=246$ GeV).  This bound is a bit difficult to interpret, because the
unitarity of gauge boson scattering requires in any case that there be new
physics, such as a Higgs boson, with a mass lighter than this.  This paper
re-examines the Appelquist-Chanowitz bound in order to clarify its meaning.
This work uses toy models with a singlet Higgs boson to unitarize gauge boson
scattering, and considers other possibilities for the new physics affecting
the fermion mass.  This new physics has the effect of changing the Higgs
boson--fermion--anti-fermion coupling.  New physics cannot significantly alter
this coupling unless it is substantially lighter than the Appelquist-Chanowitz
bound.

\bigskip

\Date{7/94}

\vfil\eject

\newsec{Introduction}

Several years ago Appelquist and Chanowitz pointed out that there is a bound
on the scale of fermion mass generation \ref\AC{T. Appelquist and M. S.
Chanowitz, \prl{59}{2405}{1987}} imposed by the unitarity of the process $f
{\bar f} \to V_L V_L$, where $f$ is a massive fermion and $V_L$ is a
longitudinal $W^\pm$ or $Z$.  The argument may be summarized as follows.
Consider the calculation of this process in unitary gauge in the standard
model.  There are four tree-level diagrams.  One exchanges a fermion in the
$t$ channel; one has a fermion in the $u$ channel; one has a neutral gauge
boson, a $Z$ or $\gamma$, in the $s$ channel.  The fourth diagram has a Higgs
boson in the $s$ channel.  The Higgs boson exchange diagram, the argument
goes, is the sign of the new physics giving mass to the fermion.  Compute the
amplitude for a heavy lepton--anti-lepton pair to annihilate into a pair of
longitudinal $W$ bosons.  The process of interest is the one that does not
conserve chirality.  The calculation of the first three diagrams yields a
result which grows with $s$, the center-of-mass energy squared, like
\eqn\higheamp{
a({\bar \ell}_+ \ell_- \to W^+_L W^-_L) \approx {m_\ell \sqrt{s} \over v^2}
{}~~.
}
At high energies the Higgs boson exchange diagram makes a contribution to the
amplitude of precisely the same form but with the opposite sign.  Thus, at
energies above the mass of the Higgs boson, the growth in this amplitude is
cut off.  The requirement that the spin-zero partial wave not violate
unitarity --- that it stay smaller than one in magnitude --- implies that the
scale $M_U$ before which the calculation without the Higgs boson must break
down is bounded by\foot{For a colored particle, there is an additional factor
of $\sqrt{3}$ in the denominator.  For a top quark mass of 174 GeV, this
corresponds to a mass scale of about 10 TeV.}
\eqn\maxh{
M_U < {16 \pi v^2 \over m_\ell}
{}~~.
}
Therefore, it is argued, the mass of the Higgs boson or whatever else gives
mass to the fermion must be less than $M_U$.

The difficulty here is that there is a more stringent bound on the mass of the
Higgs boson or other new physics, coming from the unitarity of the process
$V_L V_L \to V_L V_L$.  New physics must appear before about a mass of 1 TeV
\ref\LQT{B.~W.~Lee, C.~Quigg, and H.~B.~Thacker, Phys. Rev. {\bf D16}, 1519
(1977)\semi M. S. Chanowitz and M. K. Gaillard, \npb{261}{379}{1985}}.  If the
mass of new physics is postponed to near the 1 TeV bound, the scattering of
the gauge bosons off each other makes the calculation of the three tree-level
diagrams unreliable long before the multi-TeV energies implied by \maxh\ (see
\fig\figI{Gauge boson rescattering makes the tree-level computation of gauge
boson production unreliable at a low energy scale.}).  What then is the
significance of the unitarity constraint?

Since we know that there must be a symmetry breaking sector at 1 TeV or below,
the outstanding questions are really a bit more subtle.  Does this symmetry
breaking sector have to couple to the fermions, and if so, how?  At what scale
is the physics of flavor?  If one believed, for instance, that the one-doublet
standard model were the theory of everything, then the physics of flavor lives
at the Planck scale.  Unitarity implies nothing.  On the other hand, there are
possibilities for putting the physics of flavor at lower scales.  One
interesting example is the top-mode standard model
\ref\topmode{V. Miranski, M. Tanabashi and K. Yamawaki,
\plb{221}{177}{1989}\semi V. Miranski, M. Tanabashi and K. Yamawaki,
\mpla{4}{1043}{1989}\semi W. Bardeen, C. Hill, M. Lindner,
\prd{41}{1647}{1990}.}, which can be used to put the physics of fermion masses
at essentially any scale one wishes.

In this note, for simplicity, it is assumed that the gauge boson rescattering
is unitarized by the addition of a single Higgs boson.  This is only an
example of the kind of symmetry breaking sector one might encounter; a
multi-Higgs boson model would do just as well.  The only requirement is that
once the resonances are included, the theory must look renormalizable.
However, the common feature of all models that put the physics of fermion
masses at a high scale is that at low energies they look like a renormalizable
model that includes a fermion mass.  For example, the top-mode standard model
is almost indistinguishable at low energies from the standard model\foot{The
usual sort of technicolor model, by contrast, looks at no scale even
approximately like a renormalizable model of resonances.  Thus the
considerations of this paper are not really applicable to technicolor.  This
point is discussed in the conclusions.}.

No matter what the symmetry breaking sector is, gauge invariance dictates that
there be no mass of the fermion in the absence of a vacuum expectation value.
Therefore, the fermions will always get their mass from the Higgs boson and
there will always be a Higgs boson--fermion--anti-fermion coupling.  However,
the coupling strength may not have its standard model value.  The question
this paper asks is What scale must the new physics have in order to alter the
the Higgs boson--fermion--anti-fermion coupling substantially, by order one?
This paper examines the mass scale of the new physics in two kinds
of effective field theories.  In the first class, the new physics gets its
mass by coupling to the vacuum expectation value, and treatment using a
non-linear Lagrangian is appropriate.  The second category includes those
models in which the mass of the new physics is introduced in an $SU(2)_W
\times U(1)_Y$ invariant fashion.

\newsec{Non-Linear Models}

Consider a non-linear chiral Lagrangian \ref\CCWZ{S. Weinberg, \pr {166}{1968}
{1568}\semi S.  Coleman, J. Wess, and B. Zumino, \pr {177}{2239}{1969}\semi
C. Callan, S.  Coleman, J. Wess, and B. Zumino \pr {177}{2246}{1969}.} with
gauge group $SU(2)_W \times U(1)_Y$ broken to the electromagnetic $U(1)$
\ref\chirstd{T. Appelquist, Lectures presented at the 21st Scottish
Universities Summer School in Physics, St. Andrews, Scotland, Aug 10-30, 1980.
Published in {\it Scottish Summer School 1980}, p 385\semi
A. Longhitano, \prd{22}{1166}{1980} and \npb{188}{118}{1981}\semi
R.~Renken and M.~Peskin, \npb{211}{93}{1983}\semi
M.~Golden and L.~Randall, \npb{361}{1990}{3}\semi
B.~Holdom and J.~Terning, \plb{247}{88}{1990}\semi
A.~Dobado, D.~Espriu, and M.~J.~Herrero, \plb{255}{405}{1990}\semi
H.~Georgi,\npb{363}{301}{1991}.}.  Aside from some
number of fermions whose mass is neglected, there is one massive fermion $f$,
of mass $m_f$, the left-handed component of which is the upper member of a
doublet $\psi_L$.  In the non-linear formalism one may construct arbitrary
vertices in a manner that preserves gauge invariance.  Define the field
corresponding to the ``swallowed'' Goldstone bosons, $w_a$
\eqn\sigmadef{
\Sigma = \exp{2 i w \cdot T \over v}
{}~~,
}
where $T^a$ are the $2 \times 2$ generators of $SU(2)$ normalized to 1/2, and
$v = 246$ GeV.  The object $\Sigma$ transforms under the weak $SU(2)_W$ as
$\Sigma \to U \Sigma$ and under hypercharge as $\Sigma \to \Sigma \exp(- i
\alpha T_3)$, and so the covariant derivative is
\eqn\covdef{
D_\mu \Sigma = \partial_\mu \Sigma + {i \over 2} g W_\mu \cdot T \Sigma
- {i \over 2} g' \Sigma T_3 B_\mu
{}~~.
}
The kinetic energy term of the Goldstone bosons is
\eqn\KEG{
{\cal L}_{KE} = {v^2 \over 4} \tr[(D^\mu \Sigma)^\dagger D_\mu \Sigma]
}
By making $SU(2)_W$ gauge transformations, one may set $\Sigma = 1$, yielding
unitary gauge.  In this gauge \KEG\ is a mass term for the $W$ and $Z$.
The object
\eqn\Fdef{
F_L \equiv \left(\matrix{1 & 0}\right) \Sigma^\dagger \psi_L
}
transforms as an $SU(2)_W$ singlet with a hypercharge equal to the electric
charge of $f$ \ref\PZ{R.~D.~Peccei and X.~Zhang, \npb{337}{269}{1990}.}.
(Here and below, replace $(1~0)$ with $(0~1)$ if the fermion is the lower
member of $\psi$.) Using this field it is possible to construct an $SU(2)_W
\times U(1)_Y$ invariant term
\eqn\Lmass{
{\cal L}_{m} = m_f {\bar f}_R F_L + h.c.
{}~~.
}
In unitary gauge this is a mass term for the quark, and it generates no
other Feynman vertices.

The calculation of Appelquist and Chanowitz probes an effective theory in
which the Lagrangian consists of \KEG, \Lmass, and the kinetic energy terms of
the gauge bosons and fermions.  While there is no way to tell definitely the
scale at which this Lagrangian breaks down, there are good arguments that it
is not much bigger than $4 \pi v$ \ref\WG{S. Weinberg, \physica{96A}{327}
{1979}\semi {\it see also} H. Georgi and A. Manohar, \npb {234}{189}{1984}.}.
The breakdown occurs first in the gauge boson scattering
process, but there is no meaningful way to use this Lagrangian for fermion
scattering above the cutoff.

To go beyond $4 \pi v$, one has to cure the unitarity problem in gauge boson
scattering.  Any renormalizable model with spontaneous symmetry breaking will
work, but for the purposes here it is sufficient to consider the simplest
possible case, the one-doublet standard model, with a single Higgs boson.
Include a new singlet field $H$.  Couplings of the Higgs boson to the
Goldstone bosons are added to the Lagrangian.  Define
\eqn\Phidef{
\Phi \equiv {H + v \over \sqrt{2}} \Sigma
{}~~.
}
If gauge boson scattering is to be unitary at high energies, the couplings of
the Higgs boson to the other particles must be such that the Goldstone boson
kinetic energy term and the mass term of the Higgs boson may be written to
look like the ordinary standard model
\eqn\LnewII{
{\cal L}_U = {1 \over 2} \tr [(D^\mu \Phi)^\dagger D_\mu \Phi]
+ {m_H^2 \over 4 v^2} \left({1\over 2} \tr [\Phi^\dagger \Phi]
   - {v^2 \over 2}\right)^2
+ m_f {\bar f}_R F_L + h.c. + \ldots
{}~~,
}
where $\ldots$ represents the kinetic energy terms of the gauge bosons and
fermions.  If the mass terms of the fermion are ignored and the mass of the
Higgs boson is not too large --- less than a few hundred GeV or so --- the
gauge boson scattering will be unitary up to high energies.

This model looks almost renormalizable, but it isn't quite, because of the
absence of the coupling of the massive fermion to the symmetry breaking
sector, the Higgs boson.  The standard model mass term for the fermion is
\eqn\mstd{
{\cal L}_{mf} = m_f \left(1 + {H \over v}\right) {\bar f}_r F_L + h.c.
{}~~
}
which contains such a coupling.  In fact, in the absence of the $H {\bar f}_R
F_L$ vertex the Lagrangian doesn't even make sense as an effective theory,
because diagrams such as the one shown in \fig\figII{One of the diagrams that
are infinite in an $R_\xi$ gauge and contribute to the $H {\bar f}_R F_L$
coupling.} are infinite in an $R_\xi$ gauge and contribute to the $H {\bar
f}_R F_L$ coupling.  Evaluating these diagrams in $4-\epsilon$ dimensions, one
finds a divergence\foot{Note that there is no divergence proportional to $g^2
m_f$ or $g'^2 m_f$.}
\eqn\diver{
{m_f^3 \over 4 \pi^2 v^3}
\left({2 \over \epsilon} + \log(\mu^2) + \hbox{finite}\right)
H {\bar f}_R F_L + h.c.
}
Because of the existence of the infinity, a counterterm of this form must be
added.  In the effective theory the renormalized value of this coupling may be
chosen to be anything at all, it does not have to satisfy the relationship in
\mstd.

Suppose the renormalized coupling between the fermion and the symmetry
breaking sector is negligibly small compared to the value it should have had
in order to make the theory renormalizable.  In our sample case, this means
that we take the renormalized Higgs boson--fermion--anti-fermion coupling to
be much smaller than given by \mstd.  The non-renormalizable theory has a
mismatch between this coupling and the fermion mass, and therefore, new physics
will eventually be required at some scale.  It is possible to see what that
scale is by considering loop diagrams such as the one shown in
\fig\figIII{An infinite loop diagram requiring a ${\bar f}_R F_L {\bar f}_R
F_L$ counterterm.}.  This diagram is infinite; the infinity is
\eqn\ttttdiver{
{3 m_f^2 \over 256\pi^2 v^4}
\left({2 \over \epsilon} + \log(\mu^2) + \hbox{finite}\right)
{\bar f}_R F_L {\bar f}_R F_L
}
In the standard model this infinity is cancelled by a diagram of the same
form, but with one of the Goldstone bosons replaced a Higgs boson.  Since the
relationship between the various couplings has been destroyed in this model,
the infinity is really present, and a counterterm must be added.  In the
absence of some sort of fine tuning, it would be surprising if the size of the
renormalized counterterm were much smaller than $(3 m_f^2)/(256\pi^2 v^4)$.
This implies a new mass scale has entered the problem, the scale that
suppresses the four-fermion operator: $16 \pi v^2/\sqrt{3} m_f$.  The
existence of this four-fermion operator implies that there must be new physics
at or below that scale.

Alternatively, one may see the new scale in the way that Appelquist and
Chanowitz did, by finding where $f {\bar f} \to V_L V_L$ unitarity is
destroyed.  In this model, if the Higgs boson mass is light, the calculation
of Appelquist and Chanowitz is valid because gauge boson rescattering in the
final state is negligible.  It is therefore not a coincidence that the
Appelquist-Chanowitz bound is the same as the scale of new physics in this
scenario.  Note that only in the case of a {\it light} Higgs boson is the
Appelquist-Chanowitz calculation directly valid at high energies.

However, it is probably not possible to saturate the Appelquist-Chanowitz
bound even when there is a light Higgs boson\foot{I would like to thank Howard
Georgi for pointing this out.}.  The analysis of this section applies only
when the new physics gets its mass from coupling to Higgs vacuum expectation
value (VEV).  If all or most of the mass of the new physics is fundamental,
{\it i.e.} independent of the symmetry breaking VEV, the effects of the new
physics are decoupling.  In such a case all the non-renormalizable operators
introduced by the new physics will be suppressed by inverse powers of the mass
of the new physics \ref\AC{T. Appelquist and J. Carazzone,\prd{11}{2856}
{1975}.}.  Such models can be treated by the techniques of the next section.
On the other hand, in the case under consideration in this section, the new
physics gets its mass by coupling to the Higgs VEV, and in that even in the
limit of ultrastrong coupling it is probably not possible to make it much
heavier than a few TeV.  Thus the Appelquist-Chanowitz bound is probably
irrelevant for fermions other than the top.

\newsec{Linear Models}

Consider a scenario in which the low-energy effective theory looks
approximately renormalizable, but there is high-mass physics that gets its
mass in a $SU(2)_W \times U(1)_Y$ invariant manner.  In the low-energy
effective theory applicable below the new physics, there will be gauge
invariant operators of dimension higher than four.  These will be suppressed
by powers of the large mass scale.  This sort of theory is approximately
renormalizable in the sense that one can work to any given order in inverse
powers of the large scale.

In such a situation, since the new physics has nothing to do with the
electroweak symmetry breaking, the operators in the effective theory should be
written in terms of the linearly realized Higgs sector fields.  Once again,
the one-doublet model is used as an example of what might happen in
other models; it would be easy to extend the analysis to any given symmetry
breaking sector.

It is easy to see that one may write operators that reduce or remove the Higgs
boson--fermion--anti-fermion couplings.  For example, consider the addition to
the Lagrangian of a term
\eqn\oprem{
{1 \over M_X^2} f_R \left(\matrix{1 & 0}\right)
\Phi^\dagger\Phi\Phi^\dagger \psi_L
{}~~,
}
where $M_X$ is a mass.  This dimension-six operator can be written in unitary
gauge
\eqn\opremII{
{v^3 \over M_X^2} f_R f_L \left(1 + {H \over v}\right)^3
{}~~.
}
This operator affects both the fermion's mass and its coupling to the Higgs
boson.  Suppose this operator appears in the Lagrangian with the ordinary
fermion mass term in the combination
\eqn\combi{
{3\over 2} m_f f_R f_L \left(1+{H \over v}\right)
- {1 \over 2} m_f f_R f_L \left(1 + {H \over v}\right)^3
{}~~.
}
In \combi\ there is no coupling of a single Higgs boson to the fermion --- we
have arranged a delicate cancellation.  In this case we identify
\eqn\MX{
M_X^2 = {2 v^3 \over m_f}
}

The significance of this is as follows.  One does not expect that new physics
will entirely remove the coupling of the Higgs boson to the fermions as in
\combi.  On the other hand, if there is to be a substantial alteration of the
Higgs boson--fermion--anti-fermion coupling, there must be operators like
\oprem\ suppressed by $M_X$ approximately as large as the value given in
\MX.  For fermions with mass less than $v$, $M_X$ is considerably smaller than
the scale of new physics of the previous section.  There is no $16\pi$ in this
expression, and moreover the scale $M_X$ decreases only like the square-root
of the mass of the fermion.

In fact, the mass $M$ of the new particles in this scenario is likely to be
even smaller than $M_X$.  If one considers how an operator like \oprem\ would
actually arise when particles of mass $M$ are integrated out, one expects that
there would be coupling constants in the numerator of the suppression factor.
Moreover, if the new physics doesn't create \oprem\ at tree level, then there
will be loop integral factors of $16 \pi^2$ in the denominator.  In realistic
models, therefore, it is probably impossible to substantially affect the size
of the Higgs boson--fermion coupling (at least for the top quark) in any model
in which the new physics gets its mass in an $SU(2)_W \times U(1)_Y$ invariant
fashion.

\newsec{Conclusions}

The Appelquist--Chanowitz unitarity bound on the physics giving mass to the
fermions is a bit difficult to interpret.  In a general model, it is not
possible to ignore the rescattering effects in the production of longitudinal
gauge bosons.  The existence of these effects implies that there must be a
Higgs boson or other physics below about 1 TeV.  The possibility considered in
this paper is that once the Higgs resonance is included, the model looks
approximately like a renormalizable theory with a massive fermion, but lacking
the right coupling between the fermion and the symmetry breaking sector.

The application of the considerations in this paper to conventional
technicolor \ref\techni{S.  Weinberg, \prd {19}{1277}{1979}\semi L. Susskind,
\prd {20}{2619}{1979}.} is not entirely straightforward.  Above the
scale of the condensate, treatment of the longitudinal gauge bosons is subtle,
and the calculation that includes them as fundamental particles is invalid.
Just as one cannot compute the production of the two-pion exclusive final
state in an $e^+e^-$ machine of 20 GeV center-of-mass energy, one cannot
perform the computation of $f \bar f$ to longitudinal gauge bosons in a
technicolor theory at 50 TeV.

In the usual sort of technicolor models, fermion masses come about because the
theory has non-renormalizable four-fermion operators.  The strength of the
four-fermion operators dictates the breakdown of this theory, not the
unitarity of $f {\bar f} \to V V$.

In this paper, the one-doublet standard model was considered as an example of
the kind of physics that could unitarize the gauge boson scattering.  The
Higgs boson gives mass to the fermions, just as it does in the standard model,
and there is always a Higgs boson--fermion--anti-fermion coupling.  What can
be altered, however, is the coupling of the symmetry breaking sector, the
Higgs boson, to the fermions.  Sensibility of the effective field theory
implies that the mass of the new physics is no more than a few TeV if the
coupling is to be substantially different from its standard model value.  If
the new physics gets it mass in an $SU(2)_W \times U(1)_Y$ invariant fashion,
the new physics must come in at a much lower scale, in the case of the top
quark, near the weak scale or lower.

\newsec{Acknowledgments}

I would like to thank R. Sekhar Chivukula, Mike Dugan, Howard Georgi, Ken
Lane, and Raman Sundrum for useful conversations and Mike Chanowitz and Tom
Appelquist for reading the manuscript.  This work was supported under National
Science Foundation grant number NSF PHY-9218167, and by an NSF National Young
Investigator Award.

\listrefs
\listfigs

\inpic

\newbox\phru
\setbox\phru=\hbox{\beginpicture
\setcoordinatesystem units <1truein, 1truein>
\setquadratic
\plot
0 0
0.025  0.03
0.05   0
0.075 -0.03
0.10   0
/
\endpicture}

\newbox\phrd
\setbox\phrd=\hbox{\beginpicture
\setcoordinatesystem units <1truein, 1truein>
%\stpltsmbl
\setquadratic
\plot
0 0
0.025 -0.03
0.05   0
0.075  0.03
0.10   0
/
\endpicture}

$$
\beginpicture
\setcoordinatesystem units <1truein, 1truein>

% These lines tell tex how big the picture is -- it ought to be able to
% figure this out for itself, but can't always.
\linethickness=0pt
\putrule from -1.5 0 to 1.5 0
\putrule from 0 -.5 to 0 .5
\linethickness=0.4pt

\setlinear
\plot
-1.5  0.5
-0.5  0.25
-0.5 -0.25
-1.5 -0.5
/
\arrow <.1truein> [.2,.67] from -1.5  0.5  to -1    0.375
\arrow <.1truein> [.2,.67] from -0.5  0.25 to -0.5  0
\arrow <.1truein> [.2,.67] from -0.5 -0.25 to -1   -0.375
\put {$f$} at -1.4 0.32
\put {$\bar f$} at -1.4 -0.32

\circulararc 360 degrees from .8 0 center at .5 0

\setlinear
\plot
0.2879 -0.2121
0.7121  0.2121
/
\plot
0.2 0
0.5 0.3
/
\plot
0.5 -0.3
0.8  0
/

\multiput {\copy\phrd} at -0.5  0.25 *7 .10 0 /
\multiput {\copy\phru} at -0.5 -0.25 *7 .10 0 /
\multiput {\copy\phru} at 0.7  0.25 *7 .10 0 /
\multiput {\copy\phrd} at 0.7 -0.25 *7 .10 0 /
\put {$V_L$} at 1.4  0.4
\put {$V_L$} at 1.4 -0.4

\endpicture
$$

\vfil\eject

$$
\beginpicture
\setcoordinatesystem units <1truein, 1truein>

% These lines tell tex how big the picture is -- it ought to be able to
% figure this out for itself, but can't always.
\linethickness=0pt
\putrule from -1.5 0 to 1.5 0
\putrule from 0 -.5 to 0 .5
\linethickness=0.4pt

\setlinear
\plot
-1.5  0.5
-0.5  0.25
-0.5 -0.25
-1.5 -0.5
/
\arrow <.1truein> [.2,.67] from -1.5  0.5  to -1    0.375
\arrow <.1truein> [.2,.67] from -0.5  0.25 to -0.5  0
\arrow <.1truein> [.2,.67] from -0.5 -0.25 to -1   -0.375
\put {$f$} at -1.4 0.32
\put {$\bar f$} at -1.4 -0.32

\multiput {$\cdot$} at -0.5  0.25 *10 .10 -0.025 /
\multiput {$\cdot$} at -0.5 -0.25 *10 .10  0.025 /
\put {$w^0$} at 0  0.32
\put {$w^0$} at 0 -0.32

\putrule from .5 0 to 1.5 0
\put {$H$} at 1.4 .15

\endpicture
$$

\vfil\eject

$$
\beginpicture
\setcoordinatesystem units <1truein, 1truein>

% These lines tell tex how big the picture is -- it ought to be able to
% figure this out for itself, but can't always.
\linethickness=0pt
\putrule from -1.5 0 to 1.5 0
\putrule from 0 -.5 to 0 .5
\linethickness=0.4pt

\setlinear
\plot
-1.5  0.5
-0.5  0
-1.5 -0.5
/
\arrow <.1truein> [.2,.67] from -1.5  0.5  to -1  0.25
\arrow <.1truein> [.2,.67] from -0.5  0    to -1 -0.25
\put {$f$} at -1.4 0.32
\put {$\bar f$} at -1.4 -0.32

\setlinear
\plot
 1.5  0.5
 0.5  0
 1.5 -0.5
/
\arrow <.1truein> [.2,.67] from  1.5  0.5 to  1  0.25
\arrow <.1truein> [.2,.67] from  0.5  0   to  1 -0.25
\put {$f$} at  1.4 0.32
\put {$\bar f$} at  1.4 -0.32

\put {$\cdot$} at -0.500 0.000
\put {$\cdot$} at -0.423 0.040
\put {$\cdot$} at -0.342 0.074
\put {$\cdot$} at -0.259 0.100
\put {$\cdot$} at -0.174 0.119
\put {$\cdot$} at -0.087 0.130
\put {$\cdot$} at 0.000 0.134
\put {$\cdot$} at 0.087 0.130
\put {$\cdot$} at 0.174 0.119
\put {$\cdot$} at 0.259 0.100
\put {$\cdot$} at 0.342 0.074
\put {$\cdot$} at 0.423 0.040
\put {$\cdot$} at 0.500 0.000
\put {$\cdot$} at -0.500 0.000
\put {$\cdot$} at -0.423 -0.040
\put {$\cdot$} at -0.342 -0.074
\put {$\cdot$} at -0.259 -0.100
\put {$\cdot$} at -0.174 -0.119
\put {$\cdot$} at -0.087 -0.130
\put {$\cdot$} at 0.000 -0.134
\put {$\cdot$} at 0.087 -0.130
\put {$\cdot$} at 0.174 -0.119
\put {$\cdot$} at 0.259 -0.100
\put {$\cdot$} at 0.342 -0.074
\put {$\cdot$} at 0.423 -0.040
\put {$\cdot$} at 0.500 0.000
\put {$w^0$} at 0  0.35
\put {$w^0$} at 0 -0.35

\endpicture
$$

\bye